\documentclass[%
 reprint,
 amsmath,amssymb,
 aip,
]{revtex4-2}

\usepackage{placeins}
\usepackage{afterpage}

\usepackage{graphicx}
\usepackage{dcolumn}
\usepackage{bm}

\begin{document}


\title{On correctly assessing the reversibility of the magnetocaloric effect from indirect measurements}


\author{R. Kiefe}
 \affiliation{Departamento de Física and CICECO, Universidade de Aveiro, 3810-193 Aveiro, Portugal}

\author{R. Almeida}%
\affiliation{ IFIMUP, Departamento de Física e Astronomia, Faculdade de Ciências, Universidade do Porto, rua do Campo Alegre s/n, 4169-007 Porto, Portugal
}%

\author{J. H. Belo}%
\affiliation{ IFIMUP, Departamento de Física e Astronomia, Faculdade de Ciências, Universidade do Porto, rua do Campo Alegre s/n, 4169-007 Porto, Portugal
}

\author{J. S. Amaral}
\affiliation{Departamento de Física and CICECO, Universidade de Aveiro, 3810-193 Aveiro, Portugal}


\date{\today}

\begin{abstract}
The adiabatic temperature change ($\Delta T_{ad}$) of a magnetic refrigerant can be indirectly estimated through field ($H$) and temperature ($T$) dependent magnetization ($M$) and specific heat ($C_p$) measurements. A direct integration approach for this estimation is frequently reported, which is an approximation to a rigorous mathematical approach. In this work, we propose an iterative method in small $H$ steps, to estimate $\Delta T_{ad}$ from indirect measurements. We show that this approach is able to reproduce the reversibility of the magnetocaloric effect, and provides a more accurate estimation of $\Delta T_{ad}$, up to 10\% when considering a detailed $M(H,T)$ and $Cp(H,T)$ dataset that reproduces the magnetothermal properties of gadolinium, a benchmark room-temperature magnetic refrigerant.

\end{abstract}

\maketitle


\section{Introduction}
The magnetocaloric effect (MCE) gives rise to a temperature change in magnetocaloric materials when exposed to a magnetic field, serving as the fundamental principle behind magnetic refrigeration. The MCE is a reversible process, and so, under adiabatic conditions, applying and removing an external magnetic field will keep the system at its starting temperature. The thermodynamics describing the MCE is well established, with over a century of research, and its discovery credited to Weiss and Piccard \cite{weiss1917,Smith2013}, in 1917. \par

Magnetic refrigeration benefits from a large MCE which will depend on the magnetocaloric material (refrigerant) used, and it largely dictates the performance of such devices \cite{Romero2012, Yu2010_review, Gsch30yrs,franco2018,smith2012}. The study of magnetic refrigeration hinges on accurately gauging the adiabatic temperature change ($\Delta T_{ad}$), either from indirect measurements (the refrigerant's magnetization and specific heat at different temperatures and external magnetic fields) or by direct measurement of $\Delta T_{ad}$. The accurate direct measurement of $\Delta T_{ad}$ is challenging, requiring dedicated equipment. As an alternative, $\Delta T_{ad}$ can be estimated using a detailed magnetization and specific heat dataset, both as a function of temperature and magnetic field.\par

\section{MCE - Thermodynamics}
The total differential of the total entropy of a magnetocaloric material can be written as \cite{tishin2003};

\begin{equation}
dS = \frac{C_{H,p}}{T} dT + \left( \frac{\partial M}{\partial T} \right)_{H,p} dH - \alpha_T Vdp,
\label{eq:entropy}
\end{equation}

$C_{H,p}$ is the heat capacity under constant magnetic field and pressure, $M$ is the magnetization of the material, and $\alpha_T$ is the bulk thermal expansion coefficient.\par

In an adiabatic-isobaric process ($dp = 0$ and $dS = 0$), we can write the infinitesimal temperature change due to the MCE as

\begin{equation}
\frac{C_{H,p}}{T} dT + \left( \frac{\partial M}{\partial T} \right)_{H,p} dH  = 0
\label{eq:mce}
\end{equation}

This mathematical description is standard, and eq. \ref{eq:mce} is frequently seen throughout literature \cite{tishin2007,Pecharsky99,franco2018}. From here, obtaining $\Delta T_{ad}$ diverges into two different branches \cite{Pecharsky99,smith2012}:

\begin{align}
\Delta T_{ad} &= -\int_{H_1}^{H_2} \frac{T}{C_p} \frac{\partial M}{\partial T} dH   \label{eq:integral}\\
\Delta T_{ad} &\approx -\frac{T}{C_{p,H}} \Delta S_M (T)_{\Delta H}  \label{eq:integral2}
\end{align}

Yet, eq. \ref{eq:mce} is a total differential equation, which cannot be rigorously solved with neither eq. \ref{eq:integral} nor \ref{eq:integral2}, as $\frac{\partial M}{\partial T}$ and $C_p$ are both functions of temperature and magnetic field. Reversing the limits of integration in equation \ref{eq:integral} evidently only changes the sign of $\Delta T_{ad}$, as the integral path is the same. So, the direct use of equation \ref{eq:integral} results in a non-reversible $\Delta T_{ad}$, an incorrect description of the MCE.\par

Smith \emph{et al.} have reported this inaccuracy, proposing the use of eq. \ref{eq:integral}, but ``numerically integrating in sufficiently small increments'', updating $T$ on each subsequent integral \cite{smith2012}. Also, Pecharsky and Gschneidner\cite{Pecharsky99} have criticized the use of eq. \ref{eq:integral2}, suggesting as an alternative calculating $\Delta T_{ad}$ from the isentropic difference between the $S(T)_{H_i}$ and $S(T)_{H_f}$: $\Delta T_{ad} \approx [T(S)_{H_f}-T(S)_{H_i}]_S$. \par

\subsection{Correctly assessing the reversible MCE}

Instead of successive numerical integration, as suggested by Smith \emph{et al.}, $\Delta T_{ad}$ can be estimated by approximating the total differential equation (eq. \ref{eq:entropy}), by taking small steps in magnetic field ($\delta H$), to which a small temperature change $\delta T$ is associated. This methodology is equivalent to a finite difference approach and is grounded in the accurate physical description of the MCE,

\begin{equation}
\delta T (T_i,H_i) = -\frac{T_i}{C_{p,H_i}} \frac{\partial M}{\partial T} (T_i,H_i) \delta H.
\label{eq:iterative}
\end{equation}

Then, letting the temperature evolve by iteration;

\begin{equation*}
\begin{aligned}
	T_{i+1} &= T_i + \delta T (T_i,H_i) \\
	H_{i+1} &= H_i + \delta H,
\end{aligned}
\end{equation*}

until $H_i = H_f$, where $H_f$ is the final magnetic field intensity desired. The adiabatic temperature change from this method is simply the difference between the final and initial temperatures: $\Delta T_{ad} = T_f - T_0$.\par

\begin{figure}
\centering
\includegraphics[width=.9\linewidth]{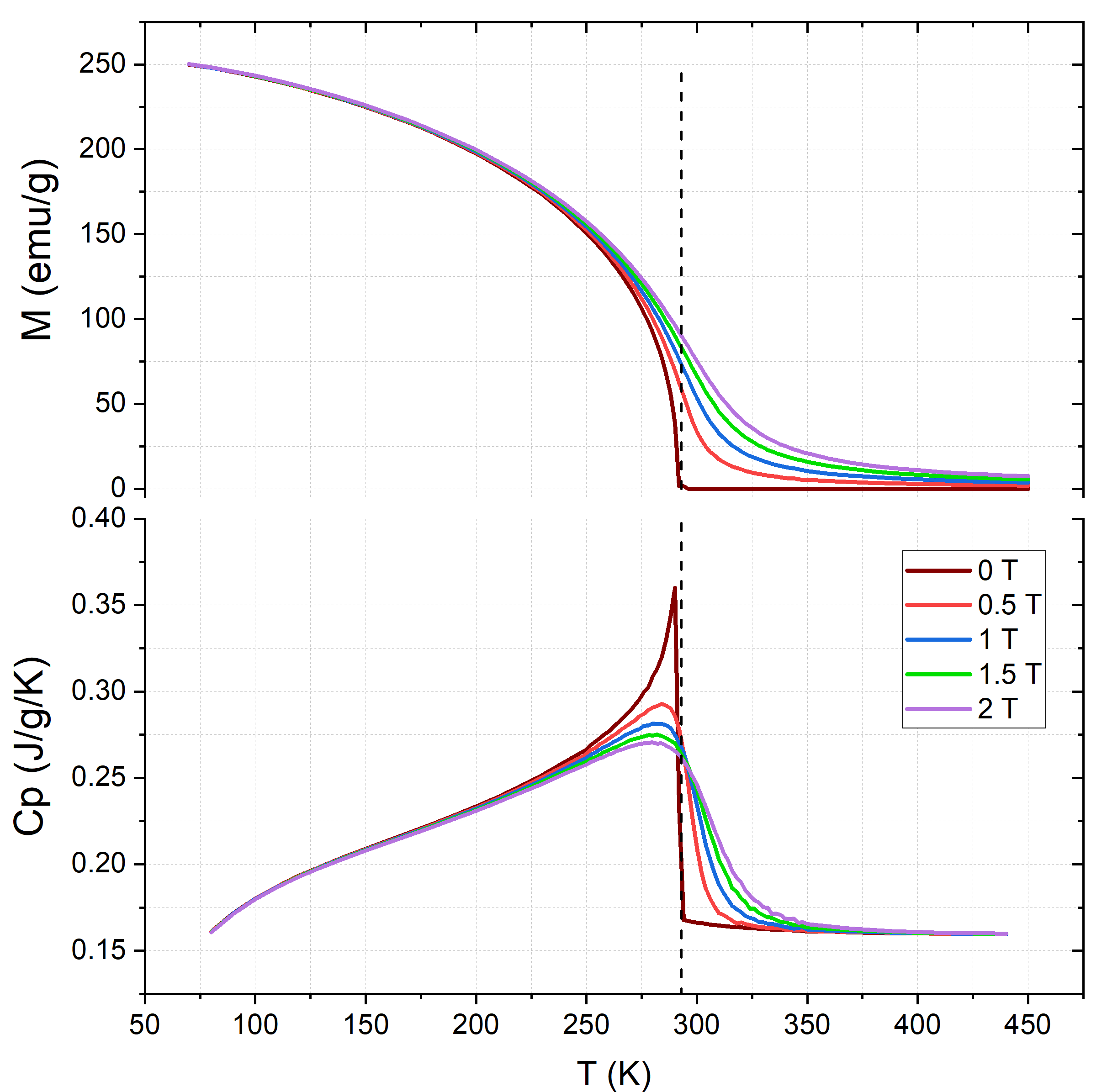}
\caption{$M(H,T)$ data (top) and $Cp(H,T)$ data (bottom) for $H = 0:0.5:2$ T, of an Ising spin 7/2 HCP lattice, where the lattice contribution to $C_p$ is obtained from the Debye model with T$_D$ = 169 K. The dashed line indicates the critical temperature for the phase transition.}
\label{fig:GdData}
\end{figure}

To obtain $\Delta T_{ad}$ from $M(H,T)$ and $C_p(H,T)$, these thermophysical properties for a magnetocaloric material are necessary and, for the iterative method (eq. \ref{eq:iterative}), its calculation requires detailed information on $M(H,T)$ and $C_p(H,T)$. In this work, we have considered detailed simulated $M(H,T)$ and $C_p(H,T)$ data that adequately replicate the thermophysical properties of gadolinium, the benchmark material for room-temperature magnetic refrigeration. These were calculated via a hexagonal close packed model lattice of spin 7/2 Ising spins, by Monte Carlo sampling of its Joint Energy and Magnetization dependent Density of States (JDOS) \cite{fss}. The nearest-neighbor magnetic exchange parameter $J$ was chosen to lead to the experimentally observed $T_c$ value of gadolinium, and the value used was $\approx$ 5.3 meV. The total specific heat is then the sum of the magnetic specific heat and the lattice contribution described by the Debye model, with a Debye temperature $T_D =$ 169 K\cite{Hill_1987}. Further details on the model, Monte Carlo methodology and comparison with experimental data are available elsewhere \cite{Almeida_2024}. Figure \ref{fig:GdData} shows $M(H,T)$ and $C_p(H,T)$ for 5 (out of 102) different external magnetic fields: $\left[0,0.5,1,1.5,2 \right]$ (T), for 79 temperature values between 80 and 440 K.\par


%

\section{Results}


Using the iterative method of eq. \ref{eq:iterative}, the adiabatic temperature change $\Delta T_{ad}$ was calculated for a field change from 0 to 2 T, using the full dataset of Fig. \ref{fig:GdData}. To verify the convergence, a set of field steps were considered, $\delta H$: $\left[ 0.01,0.05,0.025,0,001\right] $(T). By comparing each $\Delta T_{ad}$ associated to a $\delta$H, with the $\Delta T_{ad}$ obtained using the smallest field step considered ($\delta H$=0.001 T), a maximum relative difference $<0.1\%$ was observed. Also, the $\Delta T_{ad}$ from field application and removal describes a reversible process, within a maximum error of $< 0.2 \%$. \par


\begin{figure}[h]
\centering
\includegraphics[width=.8\linewidth]{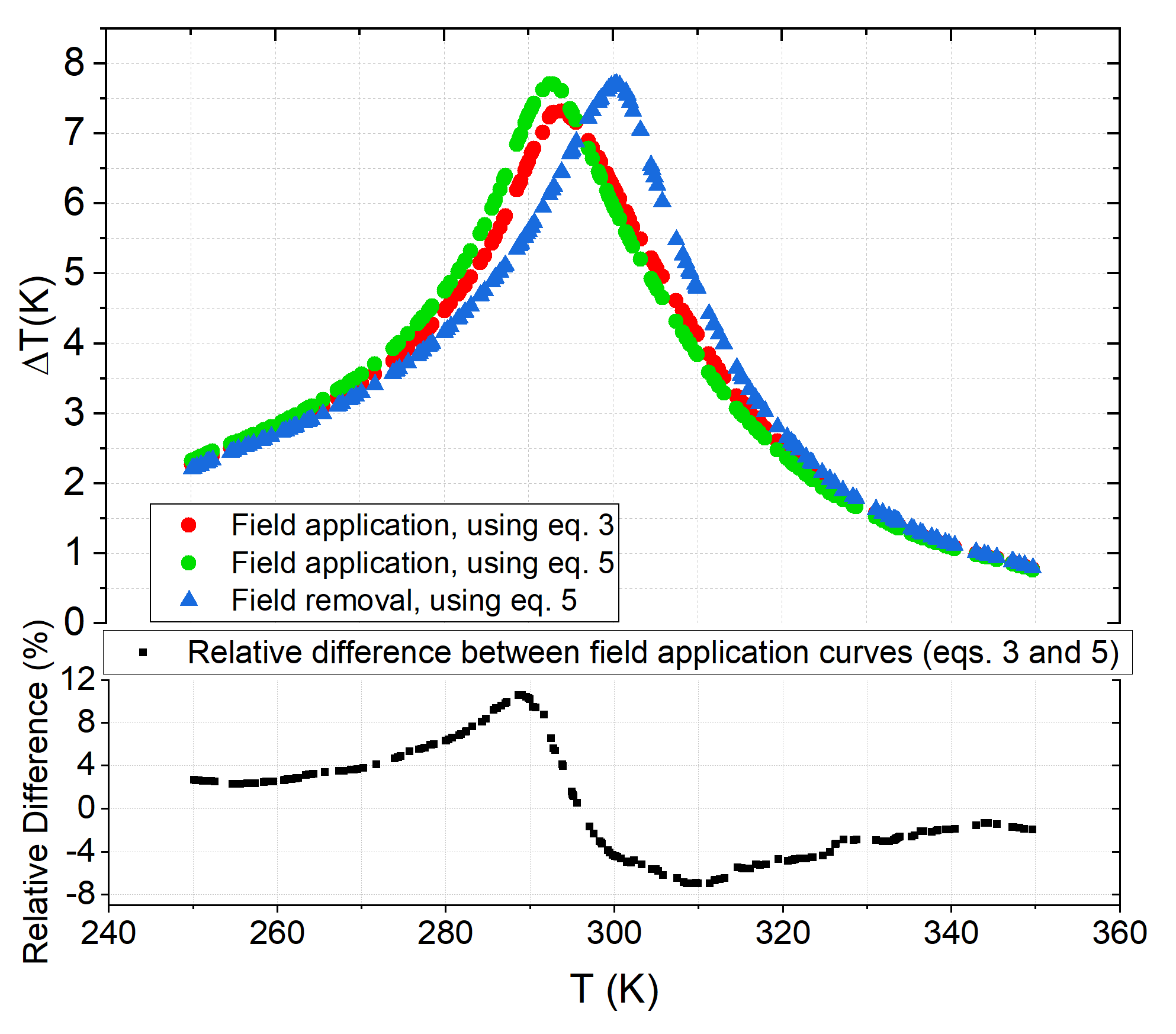}
\caption{Top - $\Delta T_{ad}$ from field application using $M(H,T)$ and $C_p(H,T)$ data from simulations that replicate Gd (Figure \ref{fig:GdData}), estimated using eq. \ref{eq:iterative} (green and blue dots) and eq. \ref{eq:integral} (red dots). Bottom - Relative difference of the $\Delta T_{ad}$ estimated from eq. \ref{eq:integral} and \ref{eq:iterative}, for field application.}
\label{fig:compare}
\end{figure}

As shown in Figure \ref{fig:compare}, the magnitude of $\Delta T_{ad}$ from field application, using the iterative method (eq. \ref{eq:iterative}) is closer to the field application estimate using eq. \ref{eq:integral}. Still, even assuming this best case scenario comparison, the difference is considerable, with a maximum relative difference larger than $10\%$. \par



\FloatBarrier
\section{Conclusion}
The conventional procedure for calculating the adiabatic temperature change ($\Delta T_{ad}$) from $M(H,T)$ and $C_p(H,T)$ data (eq. \ref{eq:integral}) is shown to not result in a reversible process, contrary to the thermodynamic description of the system. To correctly estimate the adiabatic temperature change, indirectly from $M(H,T)$ and $Cp(H,T)$ data, preserving the reversibility of the process, we propose approximating the total differential equation by an iterative method of using small steps in magnetic field (eq. \ref{eq:iterative}).\par


From detailed $M(H,T)$ and $Cp(H,T)$ data, with comparable magnetothermoal properties to gadolinium, we show that estimating $\Delta T_{ad}$ from direct integration (eq. 3) leads to a result that is 10\% deviated from the more accurate value obtained through the proposed iterative method (eq. \ref{eq:iterative}). This deviation is maximized near the Curie temperature, which is the temperature of interest, as it typically established the working temperature range for a given magnetic refrigeration device. \par

While the proposed methodology allows for the accurate indirect estimate of the $\Delta T_{ad}$ from magnetization and specific heat data, we also encourage the careful use of widely reported approximations.

\begin{acknowledgments}
This work was developed within the scope of the project CICECO-Aveiro Institute of Materials, UIDB/50011/2020, UIDP/50011/2020 \& LA/P/0006/2020, financed by national funds through the FCT/MCTES (PIDDAC), and projects PTDC/EME-TED/3099/2020, UIDP/04968/2020-Programático, UIDB/04968/2020, NECL-NORTE-010145-FEDER-022096. J.H. Belo acknowledges FCT for contract DL57/2016 reference SFRH-BPD-87430/2012 and R. Almeida acknowledges FCT for PhD. grant reference 2022.13354.BD.
\end{acknowledgments}


\section*{References}
\bibliography{apssamp.bib}

\begin{thebibliography}{13}%
\makeatletter
\providecommand \@ifxundefined [1]{%
 \@ifx{#1\undefined}
}%
\providecommand \@ifnum [1]{%
 \ifnum #1\expandafter \@firstoftwo
 \else \expandafter \@secondoftwo
 \fi
}%
\providecommand \@ifx [1]{%
 \ifx #1\expandafter \@firstoftwo
 \else \expandafter \@secondoftwo
 \fi
}%
\providecommand \natexlab [1]{#1}%
\providecommand \enquote  [1]{``#1''}%
\providecommand \bibnamefont  [1]{#1}%
\providecommand \bibfnamefont [1]{#1}%
\providecommand \citenamefont [1]{#1}%
\providecommand \href@noop [0]{\@secondoftwo}%
\providecommand \href [0]{\begingroup \@sanitize@url \@href}%
\providecommand \@href[1]{\@@startlink{#1}\@@href}%
\providecommand \@@href[1]{\endgroup#1\@@endlink}%
\providecommand \@sanitize@url [0]{\catcode `\\12\catcode `\$12\catcode
  `\&12\catcode `\#12\catcode `\^12\catcode `\_12\catcode `\%12\relax}%
\providecommand \@@startlink[1]{}%
\providecommand \@@endlink[0]{}%
\providecommand \url  [0]{\begingroup\@sanitize@url \@url }%
\providecommand \@url [1]{\endgroup\@href {#1}{\urlprefix }}%
\providecommand \urlprefix  [0]{URL }%
\providecommand \Eprint [0]{\href }%
\providecommand \doibase [0]{https://doi.org/}%
\providecommand \selectlanguage [0]{\@gobble}%
\providecommand \bibinfo  [0]{\@secondoftwo}%
\providecommand \bibfield  [0]{\@secondoftwo}%
\providecommand \translation [1]{[#1]}%
\providecommand \BibitemOpen [0]{}%
\providecommand \bibitemStop [0]{}%
\providecommand \bibitemNoStop [0]{.\EOS\space}%
\providecommand \EOS [0]{\spacefactor3000\relax}%
\providecommand \BibitemShut  [1]{\csname bibitem#1\endcsname}%
\let\auto@bib@innerbib\@empty
\bibitem [{\citenamefont {Weiss}\ and\ \citenamefont
  {Piccard}(1917)}]{weiss1917}%
  \BibitemOpen
  \bibfield  {author} {\bibinfo {author} {\bibfnamefont {P.}~\bibnamefont
  {Weiss}}\ and\ \bibinfo {author} {\bibfnamefont {A.}~\bibnamefont
  {Piccard}},\ }\bibfield  {title} {\enquote {\bibinfo {title} {{Le
  ph{\'e}nom{\`e}ne magn{\'e}tocalorique}},}\ }\href
  {https://doi.org/10.1051/jphystap:019170070010300} {\bibfield  {journal}
  {\bibinfo  {journal} {{J. Phys. Theor. Appl.}}\ }\textbf {\bibinfo {volume}
  {7}},\ \bibinfo {pages} {103--109} (\bibinfo {year} {1917})}\BibitemShut
  {NoStop}%
\bibitem [{\citenamefont {Smith}(2013)}]{Smith2013}%
  \BibitemOpen
  \bibfield  {author} {\bibinfo {author} {\bibfnamefont {A.}~\bibnamefont
  {Smith}},\ }\bibfield  {title} {\enquote {\bibinfo {title} {Who discovered
  the magnetocaloric effect?}}\ }\href
  {https://doi.org/10.1140/epjh/e2013-40001-9} {\bibfield  {journal} {\bibinfo
  {journal} {The European Physical Journal H}\ }\textbf {\bibinfo {volume}
  {38}},\ \bibinfo {pages} {507--517} (\bibinfo {year} {2013})}\BibitemShut
  {NoStop}%
\bibitem [{\citenamefont {{Romero Gómez}}\ \emph {et~al.}(2013)\citenamefont
  {{Romero Gómez}}, \citenamefont {{Ferreiro Garcia}}, \citenamefont {{De
  Miguel Catoira}},\ and\ \citenamefont {{Romero Gómez}}}]{Romero2012}%
  \BibitemOpen
  \bibfield  {author} {\bibinfo {author} {\bibfnamefont {J.}~\bibnamefont
  {{Romero Gómez}}}, \bibinfo {author} {\bibfnamefont {R.}~\bibnamefont
  {{Ferreiro Garcia}}}, \bibinfo {author} {\bibfnamefont {A.}~\bibnamefont {{De
  Miguel Catoira}}},\ and\ \bibinfo {author} {\bibfnamefont {M.}~\bibnamefont
  {{Romero Gómez}}},\ }\bibfield  {title} {\enquote {\bibinfo {title}
  {Magnetocaloric effect: A review of the thermodynamic cycles in magnetic
  refrigeration},}\ }\href
  {https://doi.org/https://doi.org/10.1016/j.rser.2012.09.027} {\bibfield
  {journal} {\bibinfo  {journal} {Renewable and Sustainable Energy Reviews}\
  }\textbf {\bibinfo {volume} {17}},\ \bibinfo {pages} {74--82} (\bibinfo
  {year} {2013})}\BibitemShut {NoStop}%
\bibitem [{\citenamefont {Yu}\ \emph {et~al.}(2010)\citenamefont {Yu},
  \citenamefont {Liu}, \citenamefont {Egolf},\ and\ \citenamefont
  {Kitanovski}}]{Yu2010_review}%
  \BibitemOpen
  \bibfield  {author} {\bibinfo {author} {\bibfnamefont {B.}~\bibnamefont
  {Yu}}, \bibinfo {author} {\bibfnamefont {M.}~\bibnamefont {Liu}}, \bibinfo
  {author} {\bibfnamefont {P.~W.}\ \bibnamefont {Egolf}},\ and\ \bibinfo
  {author} {\bibfnamefont {A.}~\bibnamefont {Kitanovski}},\ }\bibfield  {title}
  {\enquote {\bibinfo {title} {A review of magnetic refrigerator and heat pump
  prototypes built before the year 2010},}\ }\href
  {https://doi.org/https://doi.org/10.1016/j.ijrefrig.2010.04.002} {\bibfield
  {journal} {\bibinfo  {journal} {International Journal of Refrigeration}\
  }\textbf {\bibinfo {volume} {33}},\ \bibinfo {pages} {1029--1060} (\bibinfo
  {year} {2010})}\BibitemShut {NoStop}%
\bibitem [{\citenamefont {Gschneidner}\ and\ \citenamefont
  {Pecharsky}(2008)}]{Gsch30yrs}%
  \BibitemOpen
  \bibfield  {author} {\bibinfo {author} {\bibfnamefont {K.~A.}\ \bibnamefont
  {Gschneidner}, \bibfnamefont {Jr.}}\ and\ \bibinfo {author} {\bibfnamefont
  {V.~K.}\ \bibnamefont {Pecharsky}},\ }\bibfield  {title} {\enquote {\bibinfo
  {title} {Thirty years of near room temperature magnetic cooling: Where we are
  today and future prospects},}\ }\href
  {https://doi.org/10.1016/j.ijrefrig.2008.01.004} {\bibfield  {journal}
  {\bibinfo  {journal} {International journal of refrigeration-revue
  internationale du froid}\ }\textbf {\bibinfo {volume} {31}},\ \bibinfo
  {pages} {945--961} (\bibinfo {year} {2008})}\BibitemShut {NoStop}%
\bibitem [{\citenamefont {Franco}\ \emph {et~al.}(2018)\citenamefont {Franco},
  \citenamefont {Blázquez}, \citenamefont {Ipus}, \citenamefont {Law},
  \citenamefont {Moreno-Ramírez},\ and\ \citenamefont {Conde}}]{franco2018}%
  \BibitemOpen
  \bibfield  {author} {\bibinfo {author} {\bibfnamefont {V.}~\bibnamefont
  {Franco}}, \bibinfo {author} {\bibfnamefont {J.}~\bibnamefont {Blázquez}},
  \bibinfo {author} {\bibfnamefont {J.}~\bibnamefont {Ipus}}, \bibinfo {author}
  {\bibfnamefont {J.}~\bibnamefont {Law}}, \bibinfo {author} {\bibfnamefont
  {L.}~\bibnamefont {Moreno-Ramírez}},\ and\ \bibinfo {author} {\bibfnamefont
  {A.}~\bibnamefont {Conde}},\ }\bibfield  {title} {\enquote {\bibinfo {title}
  {Magnetocaloric effect: From materials research to refrigeration devices},}\
  }\href {https://doi.org/https://doi.org/10.1016/j.pmatsci.2017.10.005}
  {\bibfield  {journal} {\bibinfo  {journal} {Progress in Materials Science}\
  }\textbf {\bibinfo {volume} {93}},\ \bibinfo {pages} {112--232} (\bibinfo
  {year} {2018})}\BibitemShut {NoStop}%
\bibitem [{\citenamefont {Smith}\ \emph {et~al.}(2012)\citenamefont {Smith},
  \citenamefont {Bahl}, \citenamefont {Bjørk}, \citenamefont {Engelbrecht},
  \citenamefont {Nielsen},\ and\ \citenamefont {Pryds}}]{smith2012}%
  \BibitemOpen
  \bibfield  {author} {\bibinfo {author} {\bibfnamefont {A.}~\bibnamefont
  {Smith}}, \bibinfo {author} {\bibfnamefont {C.~R.}\ \bibnamefont {Bahl}},
  \bibinfo {author} {\bibfnamefont {R.}~\bibnamefont {Bjørk}}, \bibinfo
  {author} {\bibfnamefont {K.}~\bibnamefont {Engelbrecht}}, \bibinfo {author}
  {\bibfnamefont {K.~K.}\ \bibnamefont {Nielsen}},\ and\ \bibinfo {author}
  {\bibfnamefont {N.}~\bibnamefont {Pryds}},\ }\bibfield  {title} {\enquote
  {\bibinfo {title} {Materials challenges for high performance magnetocaloric
  refrigeration devices},}\ }\href
  {https://doi.org/https://doi.org/10.1002/aenm.201200167} {\bibfield
  {journal} {\bibinfo  {journal} {Advanced Energy Materials}\ }\textbf
  {\bibinfo {volume} {2}},\ \bibinfo {pages} {1288--1318} (\bibinfo {year}
  {2012})}\BibitemShut {NoStop}%
\bibitem [{\citenamefont {Tishin}\ and\ \citenamefont
  {Spichkin}(2003)}]{tishin2003}%
  \BibitemOpen
  \bibfield  {author} {\bibinfo {author} {\bibfnamefont {A.}~\bibnamefont
  {Tishin}}\ and\ \bibinfo {author} {\bibfnamefont {Y.}~\bibnamefont
  {Spichkin}},\ }\href {https://doi.org/10.1201/9781420033373} {\emph {\bibinfo
  {title} {The Magnetocaloric Effect and its Applications}}}\ (\bibinfo
  {publisher} {CRC Press},\ \bibinfo {year} {2003})\BibitemShut {NoStop}%
\bibitem [{\citenamefont {Tishin}(2007)}]{tishin2007}%
  \BibitemOpen
  \bibfield  {author} {\bibinfo {author} {\bibfnamefont {A.}~\bibnamefont
  {Tishin}},\ }\bibfield  {title} {\enquote {\bibinfo {title} {Magnetocaloric
  effect: Current situation and future trends},}\ }\href
  {https://doi.org/https://doi.org/10.1016/j.jmmm.2007.03.015} {\bibfield
  {journal} {\bibinfo  {journal} {Journal of Magnetism and Magnetic Materials}\
  }\textbf {\bibinfo {volume} {316}},\ \bibinfo {pages} {351--357} (\bibinfo
  {year} {2007})}\BibitemShut {NoStop}%
\bibitem [{\citenamefont {Pecharsky}\ and\ \citenamefont
  {Gschneidner}(1999)}]{Pecharsky99}%
  \BibitemOpen
  \bibfield  {author} {\bibinfo {author} {\bibfnamefont {V.~K.}\ \bibnamefont
  {Pecharsky}}\ and\ \bibinfo {author} {\bibfnamefont {J.}~\bibnamefont
  {Gschneidner}, \bibfnamefont {K.~A.}},\ }\bibfield  {title} {\enquote
  {\bibinfo {title} {{Magnetocaloric effect from indirect measurements:
  Magnetization and heat capacity}},}\ }\href
  {https://doi.org/10.1063/1.370767} {\bibfield  {journal} {\bibinfo  {journal}
  {Journal of Applied Physics}\ }\textbf {\bibinfo {volume} {86}},\ \bibinfo
  {pages} {565--575} (\bibinfo {year} {1999})}\BibitemShut {NoStop}%
\bibitem [{\citenamefont {Inácio}, \citenamefont {Ferreira},\ and\
  \citenamefont {Amaral}(2022)}]{fss}%
  \BibitemOpen
  \bibfield  {author} {\bibinfo {author} {\bibfnamefont {J.~C.}\ \bibnamefont
  {Inácio}}, \bibinfo {author} {\bibfnamefont {A.~L.}\ \bibnamefont
  {Ferreira}},\ and\ \bibinfo {author} {\bibfnamefont {J.~S.}\ \bibnamefont
  {Amaral}},\ }\href@noop {} {\enquote {\bibinfo {title} {Accurate estimate of
  the joint density of states via flat scan sampling},}\ } (\bibinfo {year}
  {2022}),\ \Eprint {https://arxiv.org/abs/2203.02718} {arXiv:2203.02718
  [cond-mat.stat-mech]} \BibitemShut {NoStop}%
\bibitem [{\citenamefont {Hill}\ \emph {et~al.}(1987)\citenamefont {Hill},
  \citenamefont {Collocott}, \citenamefont {Jr},\ and\ \citenamefont
  {Schmidt}}]{Hill_1987}%
  \BibitemOpen
  \bibfield  {author} {\bibinfo {author} {\bibfnamefont {R.~W.}\ \bibnamefont
  {Hill}}, \bibinfo {author} {\bibfnamefont {S.~J.}\ \bibnamefont {Collocott}},
  \bibinfo {author} {\bibfnamefont {K.~A.~G.}\ \bibnamefont {Jr}},\ and\
  \bibinfo {author} {\bibfnamefont {F.~A.}\ \bibnamefont {Schmidt}},\
  }\bibfield  {title} {\enquote {\bibinfo {title} {The heat capacity of
  high-purity gadolinium from 0.5 to 4 k and the effects of interstitial
  impurities},}\ }\href {https://doi.org/10.1088/0305-4608/17/9/013} {\bibfield
   {journal} {\bibinfo  {journal} {Journal of Physics F: Metal Physics}\
  }\textbf {\bibinfo {volume} {17}},\ \bibinfo {pages} {1867} (\bibinfo {year}
  {1987})}\BibitemShut {NoStop}%
\bibitem [{\citenamefont {Almeida}\ \emph {et~al.}(2024)\citenamefont
  {Almeida}, \citenamefont {Freitas}, \citenamefont {Fernandes}, \citenamefont
  {Kiefe}, \citenamefont {Araújo}, \citenamefont {Amaral}, \citenamefont
  {Ventura}, \citenamefont {Belo},\ and\ \citenamefont {Silva}}]{Almeida_2024}%
  \BibitemOpen
  \bibfield  {author} {\bibinfo {author} {\bibfnamefont {R.}~\bibnamefont
  {Almeida}}, \bibinfo {author} {\bibfnamefont {S.~C.}\ \bibnamefont
  {Freitas}}, \bibinfo {author} {\bibfnamefont {C.~R.}\ \bibnamefont
  {Fernandes}}, \bibinfo {author} {\bibfnamefont {R.}~\bibnamefont {Kiefe}},
  \bibinfo {author} {\bibfnamefont {J.~P.}\ \bibnamefont {Araújo}}, \bibinfo
  {author} {\bibfnamefont {J.~S.}\ \bibnamefont {Amaral}}, \bibinfo {author}
  {\bibfnamefont {J.~O.}\ \bibnamefont {Ventura}}, \bibinfo {author}
  {\bibfnamefont {J.~H.}\ \bibnamefont {Belo}},\ and\ \bibinfo {author}
  {\bibfnamefont {D.~J.}\ \bibnamefont {Silva}},\ }\bibfield  {title} {\enquote
  {\bibinfo {title} {Rotating magnetocaloric effect in
  polycrystals—harnessing the demagnetizing effect},}\ }\href
  {https://doi.org/10.1088/2515-7655/ad1c61} {\bibfield  {journal} {\bibinfo
  {journal} {Journal of Physics: Energy}\ }\textbf {\bibinfo {volume} {6}},\
  \bibinfo {pages} {015020} (\bibinfo {year} {2024})}\BibitemShut {NoStop}%
\end{thebibliography}%

\end{document}